\documentclass{aastex61}
\newcommand{\alphair}{$\alpha_{IR}$}
\newcommand\ha{H$\alpha$}
\newcommand\stwo{[S~II]}

\shorttitle{Herbig-Haro Objects in \object{LDN~673}}
\shortauthors{Rector et al.}
\begin{document}
\bibliographystyle{plainnat}
\title{The Discovery of Herbig-Haro Objects in \object{LDN~673}}

\correspondingauthor{Travis A. Rector}
\email{tarector@alaska.edu}

\author{T.~A. Rector}
\affil{Department of Physics and Astronomy, University of Alaska Anchorage,
    Anchorage, AK 99508, USA}
%\email{rector@uaa.alaska.edu}
\nocollaboration

\author{R.~Y. Shuping}
\affil{Space Science Institute, 4750 Walnut Street, Suite 205, Boulder, CO 80301, USA}
\nocollaboration

\author{L. Prato}
\affil{Lowell Observatory, 1400 West Mars Hill Road, Flagstaff, AZ 86001, USA}
\nocollaboration

\author{H. Schweiker}
\affil{National Optical Astronomy Observatory, Tucson, AZ 85719, USA}
\affil{Kitt Peak National Observatory, National Optical Astronomy Observatory, which is operated by the Association of Universities for Research in Astronomy, Inc. (AURA) under cooperative agreement with the National Science Foundation.}
\affil{WIYN Observatory, Tucson, AZ 85719, USA}
\nocollaboration
%\email{heidis@noao.edu}



\begin{abstract}

We report the discovery of twelve faint Herbig-Haro (HH) objects in LDN~673 found using a novel color-composite imaging method that reveals faint \ha\ emission in complex environments.  Follow-up observations in \stwo\ confirmed their classification as HH objects.  Potential driving sources are identified from the {\it Spitzer} c2d Legacy Program catalog and other infrared observations.  The twelve new HH objects can be divided into three groups:  Four are likely associated with a cluster of eight YSO class I/II IR sources that lie between them; five are colinear with the T~Tauri multiple star system AS 353, and are likely driven by the same source as HH~32 and HH~332;  three are bisected by a very red source that coincides with an infrared dark cloud.  We also provide updated coordinates for the three components of HH~332. Inaccurate numbers were given for this object in the discovery paper.  The discovery of HH objects and associated driving sources in this region provides new evidence for star formation in the Aquila clouds, implying a much larger T~Tauri population in a seldom-studied region.

\end{abstract}





\section{Introduction} \label{sec:intro}

Supersonic outflows from young stellar objects (YSOs) are often associated with Herbig-Haro (HH) objects.  HH objects are produced by shocks from the collision of an outflow with material in the interstellar medium, or with previously ejected outflows, producing collisionally excited forbidden lines.  In the optical, they can be readily identified by their \ha\ and \stwo\ line emission, non-stellar appearance, and nearly colinear alignment \citep{2001ApJ...546..299B}.  Only a small amount of mass (1 to 20 M$_{\earth}$) is entrained in the outflows which form HH objects.  A broad variety of morpholgies are observed, from regular series of bow shocks to amorphous clusters of small knots.  Similar to other ionized gas structures, such as HII regions and planetary nebulae, the temperatures of HH objects range from 8000 to 12,000K with electron densities of 10$^2$ to 10$^3$ cm$^{-3}$ \citep{1999A&A...342..717B}.
% * <rector@noao.edu> 2017-09-25T02:04:20.810Z:
% 
% Refs for these values?
% 
% ^.

Despite their distinctive emission and morphological characteristics, HH objects can be difficult to find in the dense, complex environments in which YSOs form.  However, their identification is important in order to understand the evolutionary state and age of a young star-forming region.  Because the driving outflows which form HH objects arise in the first few hundred thousand years of a young star's lifetime, HH objects are markers for active, ongoing star formation.  For example, the rich abundance of HH objects in the NGC~1333 molecular cloud complex points to a microburst of coeval star formation in that region \citep{1996ApJ...473L..49B}.  More embedded regions may host even larger populations of HH objects; however, obscuration may preclude straightforward characterization of complete samples.

In this paper we report on the discovery of twelve HH objects in the region of \object{LDN 673}.  This region contains highly fractured molecular cloud material and hosts the young T~Tauri star multiple AS~353 \citep{2002AAS...201.2005W,2004AJ....127..444T}. The LDN~673 cloud demarcates the high Galactic longitude terminus of the Aquila Rift (or ``Serpens-Aquila Rift'') which stretches from roughly $l = 30\deg$ to $50\deg$.  The low galactic longitude end of the Rift contains the relatively well-studied starforming regions \object{W40}, \object[Serpens South]{Serpens South molecular cloud}, \object[Serpens Main]{Serpens Cloud}, and \object{MWC 297}. \citet{2008ApJ...673L.151G} identified dozens of protostars in the Aquila-Serpens region as part of the {\it Spitzer} Gould's Belt survey. \citet{2010A&A...518L..85B} and \citet{2010A&A...518L.106K} studied hundreds of protostars and starless cores in a similar area of the Aquila rift cloud using {\it Herschel} observations made at 70$\mu$m to 500$\mu$m. In contrast, the rest of the Aquila Rift region (including LDN~673) is notable for the small number of young stars with definitive identifications \citep[e.g.,][]{2008hsf1.book...18P} compared to the abundant raw material apparently available for the star formation process \citep{1987ApJ...322..706D}.  This dearth of young stars could simply be the result of the fact that surprisingly little research has been carried out in Aquila, particularly given that the region is relatively nearby. The distance to the Aquila Rift clouds in general---and LDN~673 in particular---is somewhat uncertain, though most estimates indicate $d < 300$~pc \citep{2008hsf1.book...18P}.  For this study we adopt a distance to LDN~673 of $200 \pm 30$~pc, as in \citet{2006ApJ...647..432R}.  That the Serpens-Aquila region appears to be undergoing vigorous star formation is suggestive of the possibility that extremely young populations of protostars may be taking shape within the dense, cold gas of the giant molecular clouds elsewhere in the Aquila Rift.  In \S 2 we describe our observations.  In \S 3 we detail our source identification process.  \S 4 describes some of the specific sources detected. We provide a brief conclusion in \S 5.

\section{Observation and Search} \label{sec:obs}

LDN~673 was observed with the MOSAIC camera on the {\it Mayall}~4-meter telescope at Kitt Peak National Observatory.  MOSAIC is an optical camera that consists of eight 2048$\times$4096 CCD detectors arranged to form a 8192$\times$8192 array with 35 to 50-pixel-wide gaps between the CCDs.  With a scale of $0\farcs26$ pixel$^{-1}$, the field of view is approximately $36\arcmin\times 36\arcmin$.  To fill in the gaps and bad columns, all observations were completed in a five-exposure dither pattern with offsets of approximately 100 pixels.  

Observations were obtained on 4~August 2014 with the Harris $B$ (MOSAIC filter k1002), Harris $V$ (k1003), ``Nearly-Mould" $I$ (k1005) and \ha\ (k1009) filters.  Five exposures each of 480~sec in $B$, 300~sec in $V$, 180~sec in $I$ and 900~sec in \ha\ were obtained.   After the outflows were first detected in \ha, follow-up observations were completed in \stwo\ (ha16, H-alpha+16nm, k1013) on 31~August 2015, roughly a year later.  Five exposures of 900~sec in \stwo\ were obtained.  The \stwo\ confirmation observations were shifted southward to search for additional outflows associated with the HH~250 group.

The data were reduced with the IRAF package MSCRED in the standard manner.  The world coordinate system (WCS) was determined via stars from the USNO-B1.0 catalog \citep{2003AJ....125..984M} with a global solution RMS of better than $0\farcs4$ in all cases. This is assumed as the accuracy for all measured positions.  All of the images were projected onto a common WCS to correct for geometric distortion.

To better see faint HH objects, the data were registered using the common WCS and then combined to form a color-composite image with the methodology described in \citet{2007AJ....133..598R}.  Specifically, in the color images, the broadband filters $BVI$ were assigned the colors blue, green, and orange respectively;  \ha\ was assigned to red.  An advantage of searching for HH objects in this manner is that the red \ha\ emission is distinct and readily visible in contrast to the other colors.  Further, the broadband filters reveal the relative amount of obscuration from dust and gas.  Thus, faint outflows can be found more easily in the complex environments typical of star-forming regions.  The \stwo\ observations were not used in the color-composite image but were obtained later for confirmation.

All of the HH objects were visible in the \ha\ and \stwo\ filters, but were not detected in the broadband $I$ filter.  Thus we are confident they are sources of line emission only.  The objects were named in order of increasing right ascension within each grouping.  
%When objects are thought to be part of the same outflow (e.g., HH~???) they share a single HH number.  When it is unclear if individual knots are part of the same flow, they are given separate HH numbers. 
Their coordinates are given in Table~\ref{tbl:newHH}.  The given positions correspond to the brightest knot in each object.  Because the sources are extended, the positions are only given to an accuracy of 1".

\begin{deluxetable}{lll}
\tablecaption{Newly Discovered HH objects in LDN~673\label{tbl:newHH}}
\tablewidth{0pt}
\tablehead{\colhead{ID} & \colhead{RA(2000)} & \colhead{DEC}}
\startdata
HH~1183 & 19:20:24.8  & +11:20:10 \\
HH~1184 & 19:20:25.1  & +11:19:54 \\
HH~1185 & 19:20:28.7  & +11:19:24 \\
HH~1186 & 19:20:29.5  & +11:19:43 \\
\hline
HH~1187 & 19:20:11.9  & +11:00:40 \\
HH~1188 & 19:20:21.3  & +11:01:16 \\
HH~1189 & 19:20:24.5  & +11:01:28 \\
HH~1190 & 19:21:10.9  & +11:03:50 \\
HH~1191 & 19:21:13.2  & +11:04:01 \\
\hline
HH~1192 & 19:21:26.0  & +11:19:23 \\
HH~1193 & 19:21:42.6  & +11:22:41 \\
HH~1194 & 19:21:43.7  & +11:22:51 \\
\enddata
\end{deluxetable}

\section{Infrared Source Identification} \label{sec:irsources}

HH objects are usually associated with YSOs that are faint or invisible at optical wavelengths, given their obscuration by natal dust clouds, but bright in the infrared as the result of warm dust in a circumstellar disk and/or envelope.   WISE and 2MASS images of the LDN~673 region reveal a number of very red sources, many of which are near or coincident with the HH objects discovered in the MOSAIC images.  The northern part of the LDN~673 clouds was also included in the {\it Spitzer Space Telescope} \citep{2004ApJS..154....1W} Cores-to-Disks (c2d) Legacy Program \citep{2003PASP..115..965E,2009ApJS..181..321E}, which included images and source extractions from both the IRAC and MIPS instruments~\citep{2004ApJS..154...10F,2004ApJS..154...25R}.  These {\it Spitzer} data were then combined with 2MASS source magnitudes to produce a full IR spectral energy distribution (SED) for each source from 1~to~70~\micron.  Using the Infrared Science Archive (IRSA)\footnote{
\url{http://irsa.ipac.caltech.edu/}} 
we identified a number of candidate YSOs from c2d source catalogs that might be driving the various clusters of HH objects.  The candidate sources are listed in Table~\ref{tbl-YSOs}. We only searched for sources near the HH objects in question, therefore this is not a complete sample of YSOs in the LDN~673 cloud.  

\begin{deluxetable}{llllllll}
\tabletypesize{\scriptsize}
\tablecaption{Candidate IR sources driving HH flows in LDN~673\label{tbl-YSOs}}
\tablewidth{0pt}
\tablehead{\colhead{Name} & \colhead{Catalog} & \colhead{ID} & \colhead{RA(2000)} & \colhead{DEC(2000)} &
 \colhead{$\alpha_{IR}$} & \colhead{c2d Object Type\tablenotemark{a}} & \colhead{YSO Class} 
}
\startdata
HH~1183-1186 IRS1	&	c2d	&	J192025.8+111959	&	19:20:25.80	&	11:19:59.13	&	1.34	&	star+dust(MP1)	&	I			\\
HH~1183-1186 IRS2	&	c2d	&	J192025.9+111954	&	19:20:25.85	&	11:19:53.71	&	0.98	&	YSOc-red	&	I			\\
HH~1183-1186 IRS3	&	c2d	&	J192026.0+112008	&	19:20:25.96	&	11:20:08.27	&	-0.68	&	YSOc	&	II			\\
HH~1183-1186 IRS4	&	c2d	&	J192026.0+111952	&	19:20:26.04	&	11:19:52.29	&	0.16	&	YSOc-star+dust(IR2)	&	I/II			\\
HH~1183-1186 IRS5	&	c2d	&	J192026.2+111949	&	19:20:26.17	&	11:19:49.21	&	-0.76	&	YSOc-red	&	II			\\
HH~1183-1186 IRS6	&	c2d	&	J192026.7+111956	&	19:20:26.72	&	11:19:55.56	&	-0.39	&	YSOc	&	II			\\
HH~1183-1186 IRS7	&	c2d	&	J192027.0+112011	&	19:20:27.04	&	11:20:11.43	&	0.45	&	YSOc	&	I			\\
HH~1183-1186 IRS8	&	c2d	&	J192027.3+111956	&	19:20:27.30	&	11:19:56.26	&	-0.86	&	star+dust(IR4)	&	II			\\
AS353 A		&	c2d	&	J192031.0+110155	&	19:20:30.99	&	11:01:54.68	&	0.11	&	star+dust(IR2)	&	I			\\
AS353 B		&	c2d	&	J192031.0+110149	&	19:20:31.03	&	11:01:49.15	&	-1.21	&	star+dust(IR2)	&	II			\\
AS353 C\tablenotemark{b}		&	c2d	&	J192031.9+110151	&	19:20:31.94	&	11:01:51.25	&	0.19	&	star+dust(IR2)	&	I		\\
HH~1192-1194 IRS	&	c2d	&	J192134.8+112123	&	19:21:34.81	&	11:21:23.23	&	0.72	&	red	&	I			\\
\enddata
\tablenotetext{a}{See \citet{2009ApJS..181..321E} for full descriptions and references: YSOc---Candidate YSO;
star+dust(BAND)---star plus dust component longward of BAND;
red---MIPS1 flux $>3$ times the nearest IRAC band flux;
rising---unclassified, rising SED}
\tablenotetext{b}{Previously unidentified source; see text in Section~\ref{sec:hh6-9} for discussion.}
\end{deluxetable}

YSOs are often identified using the slope of the SED between 2-20~\micron\ (\alphair).  The IR slope has already been calculated for the c2d sources and is included in the catalog along with an ``object type'' based on both \alphair\ and other criteria~\citep{2009ApJS..181..321E}.  We include both the IR slope and object type for the c2d sources in Table~\ref{tbl-YSOs} as well as the YSO class based on the customary slope ranges \citep{1987IAUS..115....1L,1993ApJ...406..122A}:
\begin{description}
\item[Class I] \alphair\ $ > 0$ (Embedded protostar)
\item[Class II] $-1.5 <$ \alphair\ $< 0$ (Optically revealed T~Tauri type star)
\item[Class III] \alphair\ $ < -1.5$ (Pre-Main sequence star with remnant dust disk)
\end{description}
Given the sophisticated nature of the filtering and identification algorithms~\citep{2009ApJS..181..321E}, the c2d object type should be a much more accurate reflection of the true nature of the source than the $\alpha_{ir}$-derived YSO class.  Objects identified as ``star+dust(BAND)'' have SEDs that are consistent with a reddened, main-sequence photosphere combined with strong infrared excess emission ($>3\sigma$) from BAND longwards.  These SEDs would be typical of young stars with actively accreting circumstellar disks (e.g., Class I or II). ``Red'' objects have a very high MIPS-to-IRAC flux ratio, indicating an SED dominated by warm dust, typical of embedded Class 0/I objects.  Other candidate YSOs (``YSOc'') are identified by their location on the CMD in a region where the source is least likely to be a background galaxy.  Individual sources associated with the HH objects are discussed in further detail below.  In all of the figures North is up and East is to the left.

\section{Individual Sources} \label{sec:sources}

\subsection{HH~1183--1186}

\begin{figure}[ht]
\plotone{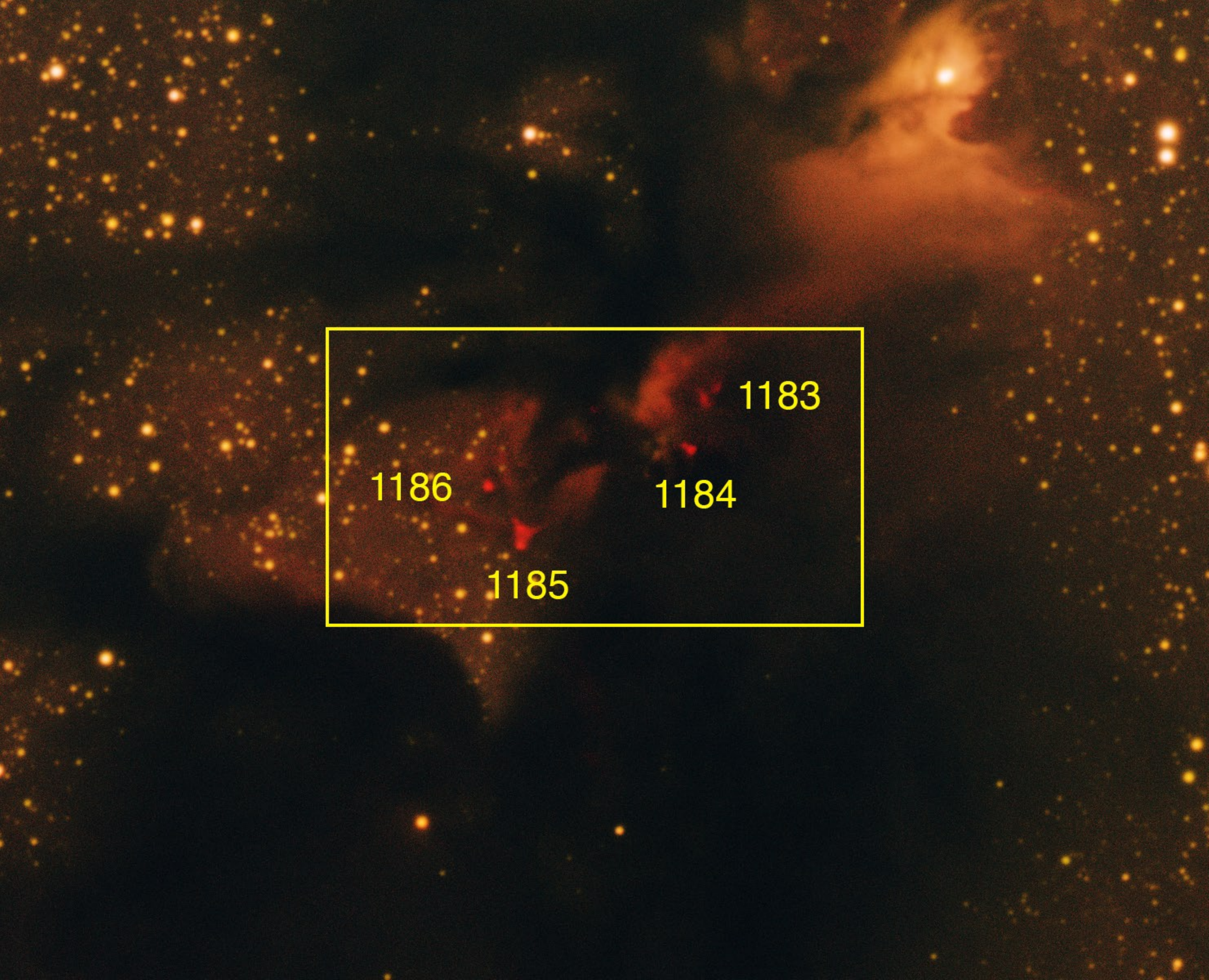}
\caption{A color composite image of HH~1183--1186.  The four objects appear to be emerging from the edge of a dense cloud. The field of view is 6\farcm4$\times$5\farcm1.  The color assignments for the filters are: $B$ (blue), $V$ (green), $I$ (orange) and \ha\ (red).  The HH objects can be distinguished by their deep red color.
\label{fig:hh1-4color}}
\end{figure}

\begin{figure}[ht]
\plottwo{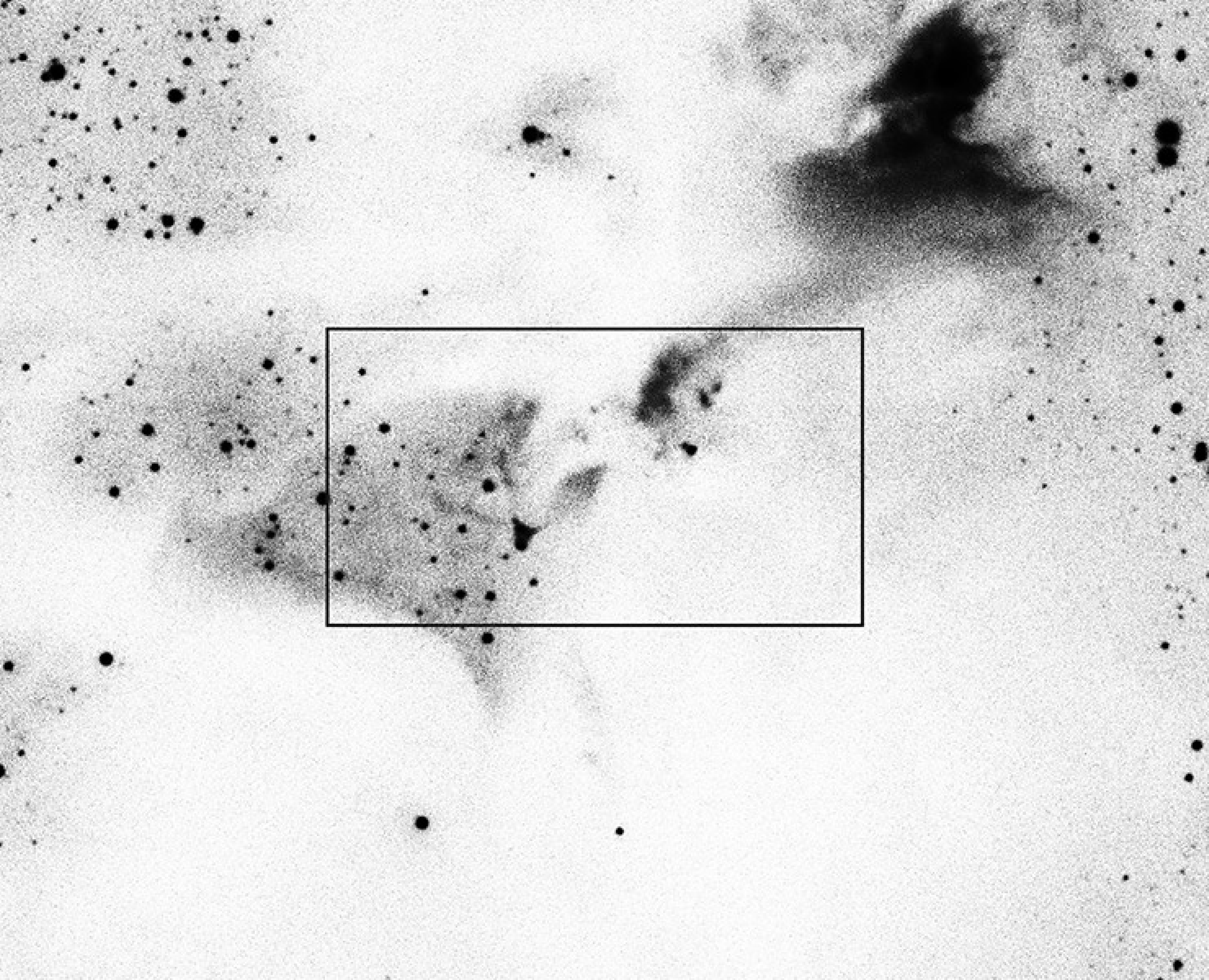}{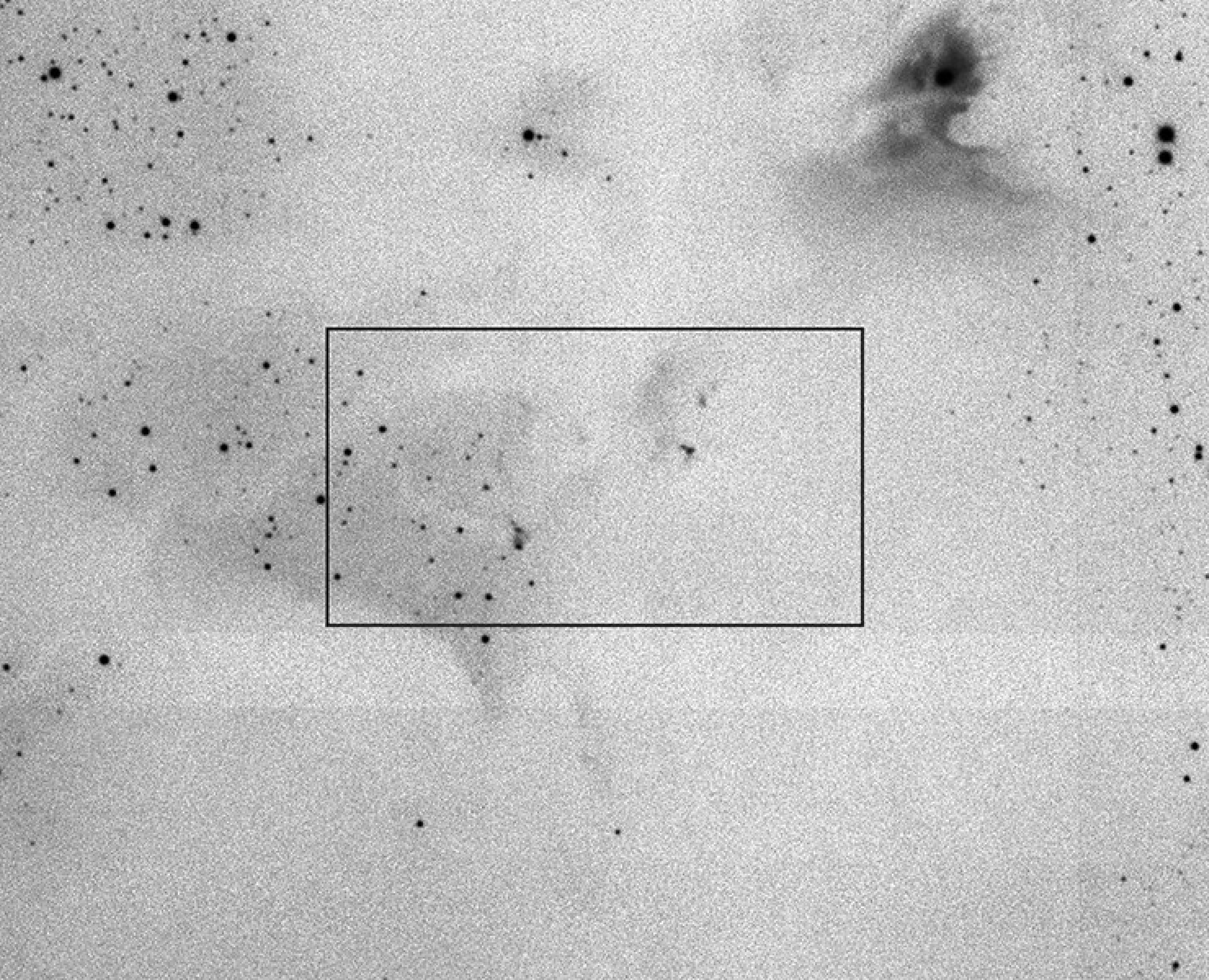}
\caption{HH~1183--1186 in \ha\ (left) and \stwo\ (right).  The field of view for both images is 5\farcm1 square.  All four objects are detected in both filters.  The box in both figures is the same region as in Figure~\ref{fig:hh1-4color}.
\label{fig:hh1-4has2}}
\end{figure}

\begin{figure}[ht]
\plottwo{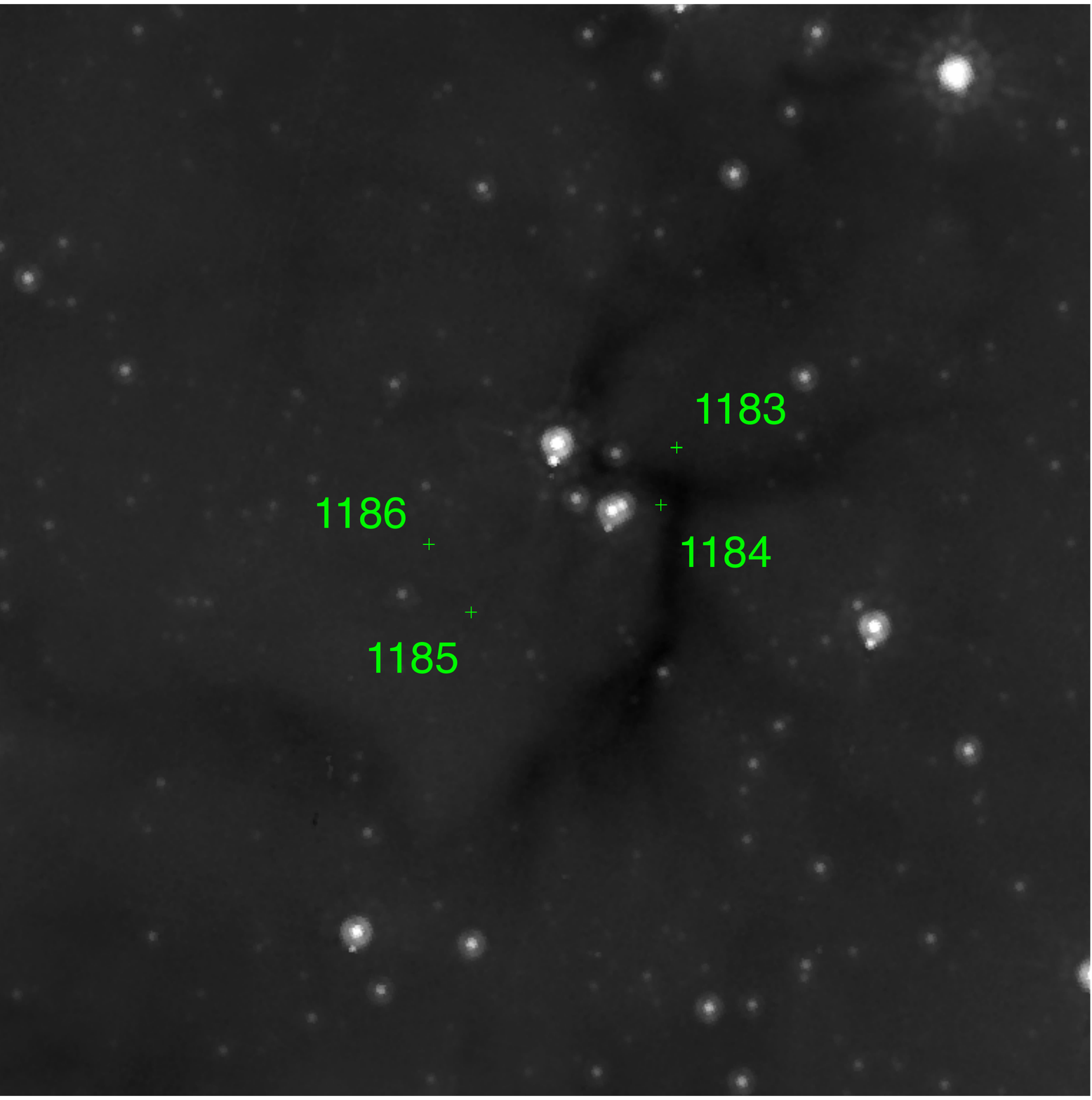}{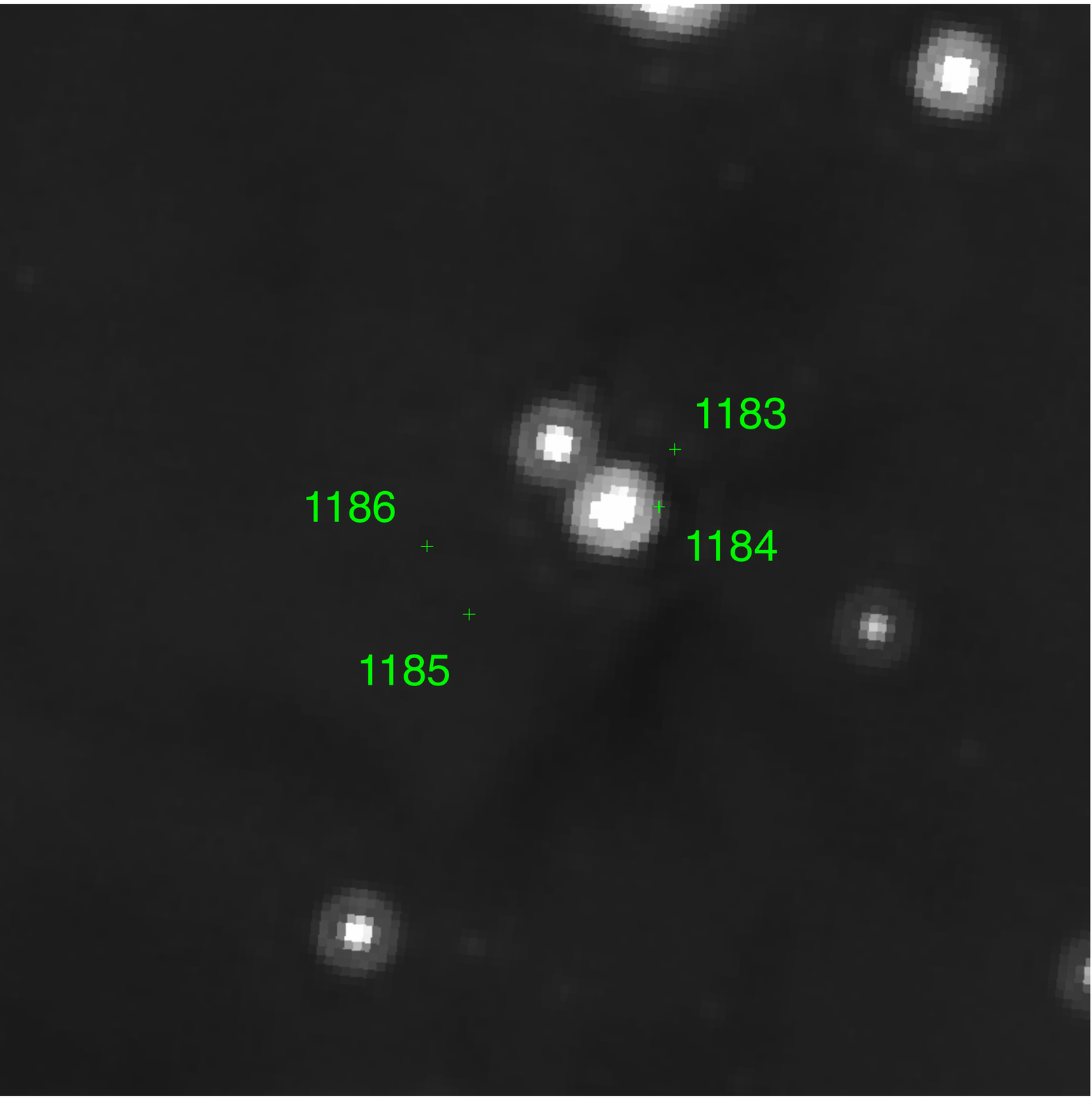}
\caption{{\it Spitzer} images of the HH~1183--1186 region using IRAC Band 4 (8 $\mu m$, left) and MIPS band 1 (24 $\mu m$, right).  The locations of the HH objects are indicated with green markers.
\label{fig:hh1-4spitzer}}
\end{figure}

\begin{figure}[ht]
\plotone{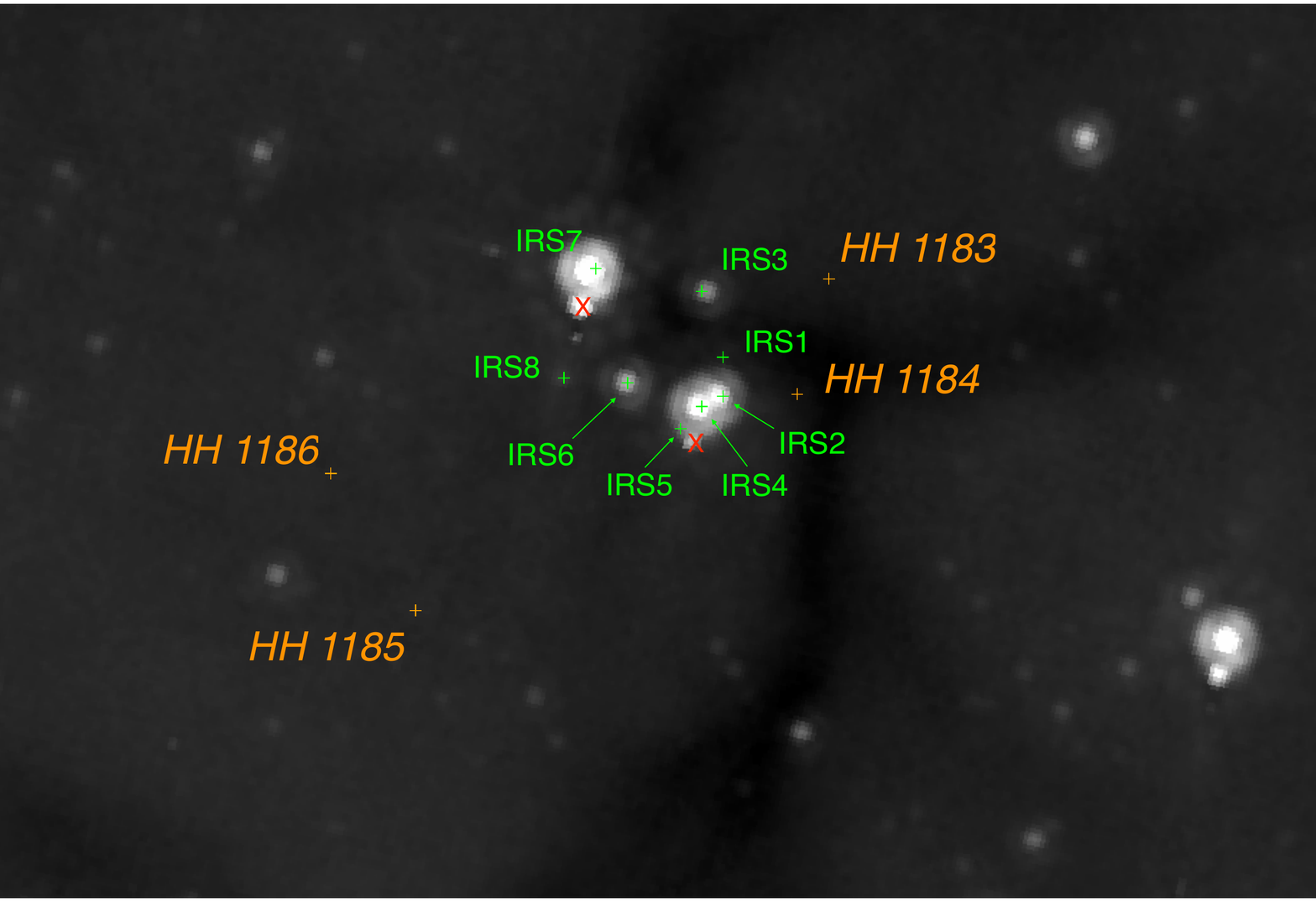}
\caption{Detail of the HH~1183--1186 region from IRAC Band 4 (8 $\mu m$) showing the known infrared sources from the c2d catalog in green, HH objects in orange, and IRAC image artifacts in red.
\label{fig:hh1-4spitzer_detail}}
\end{figure}

The color-composite image for this region is shown in Figure~\ref{fig:hh1-4color}.  The \ha\ and \stwo\ narrowband images are shown in Figure~\ref{fig:hh1-4has2}.  The optical morphologies of these sources do not suggest an obvious direction to a progenitor. However, the color optical image does suggest that the gas between HH objects is illuminated by embedded sources.
The 2MASS, WISE, and {\it Spitzer} IRAC/MIPS images (Figure~\ref{fig:hh1-4spitzer}) all reveal a small cluster of very red IR sources in between the HH objects.  A total of eight candidate Class I and II YSOs are identified in the c2d catalog (Table~\ref{tbl-YSOs} and Figure~\ref{fig:hh1-4spitzer_detail}).  HH~1183--1186~IRS2 and/or HH~1183--1186~IRS4 may be driving all four HH objects as they bisect both of the HH~1183/1185 and HH~1184/1186 pairs.  

%Not sure we need this one...Spitzer images are probably sufficient...
%\begin{figure}[ht]
% \plottwo{HH~1183-4_2MASS_square.eps}{HH~1183-4_WISE_square.eps}
% \caption{The 2MASS (left) and AllWISE (right) images of the region around HH~1--4.  The circles show the locations of the four HH objects.  The fields of view for both images are 6\farcm6 square.  For 2MASS, the image was assembled from J (1.23$\mu m$, blue), H (1.66 $\mu m$, green), and K (2.16 $\mu m$, red) filters.  For AllWISE, the image corresponds to the W1 (3.4$\mu m$, blue), W2 (12$\mu m$, green), and W4 (22$\mu m$, red) channels.
% \label{fig:hh1-4ir}}
% \end{figure}

%Need to blow this up so it matches the HH discovery image (~5' square) and label the IR sources

%\subsection{HH~250}

%HH~250:  The IRAS source associated with HH~250 breaks into 3 sources in the WISE and 2MASS catalogs.  Looks like a couple Class I sources and a ClassII/III.

%    Bo:  10 is a reflection nebula and 11 is HH 250B.
%  No additional HH objects are seen in this area.

\subsection{HH~1187--1191} \label{sec:hh6-9}

\begin{figure}[ht]
\plotone{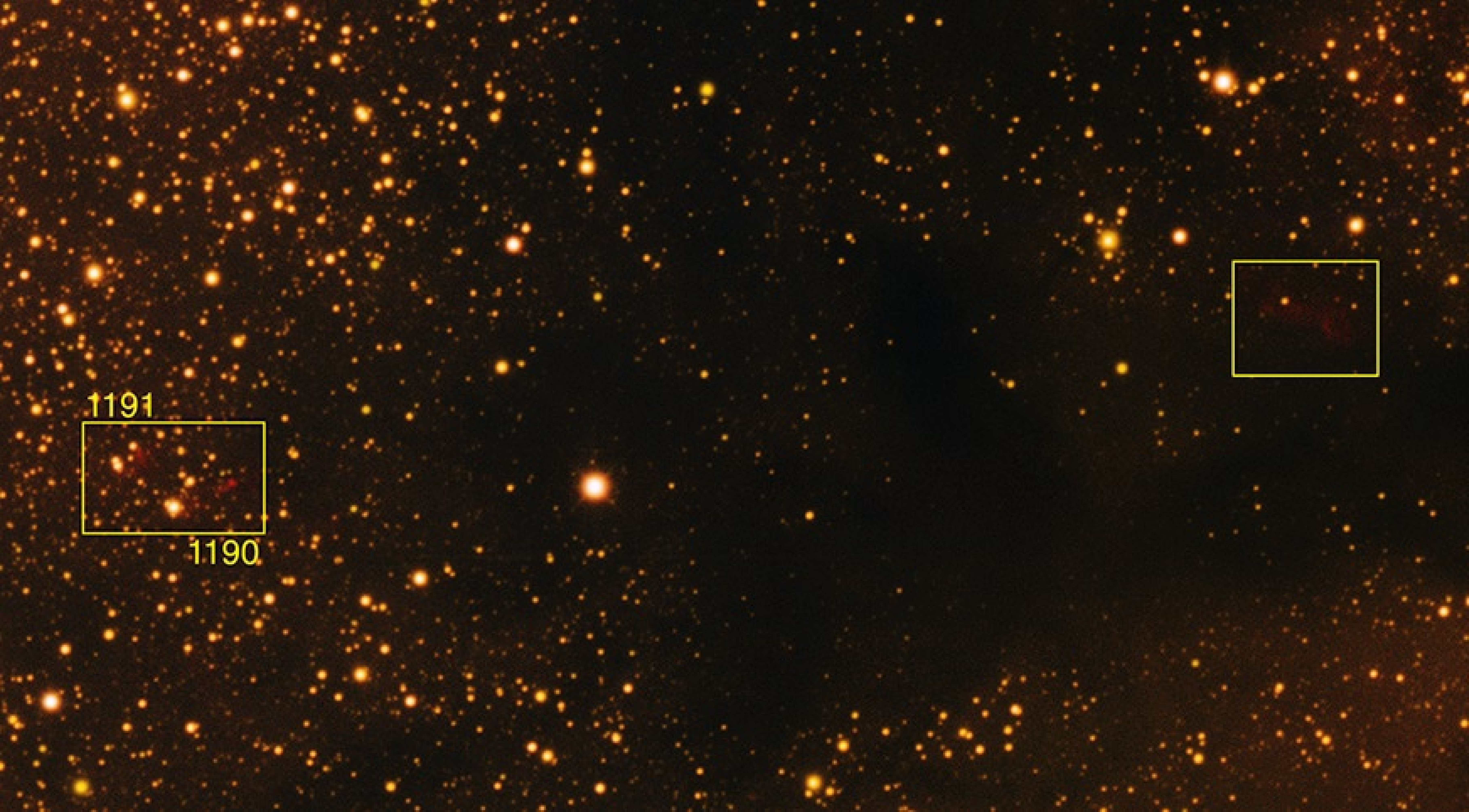}
\caption{A color composite of the region near HH~1190--1191, which are visible in the box on the left side of the image.  A faint, extended nebulosity, likely a reflection nebula from an \ha-strong source, is visible in the box on the right.  The field of view is 9\farcm0$\times$5\farcm0.  The colors are the same as for Figure~\ref{fig:hh1-4color}.
\label{fig:hh567color}}
\end{figure}

\begin{figure}[ht]
\plottwo{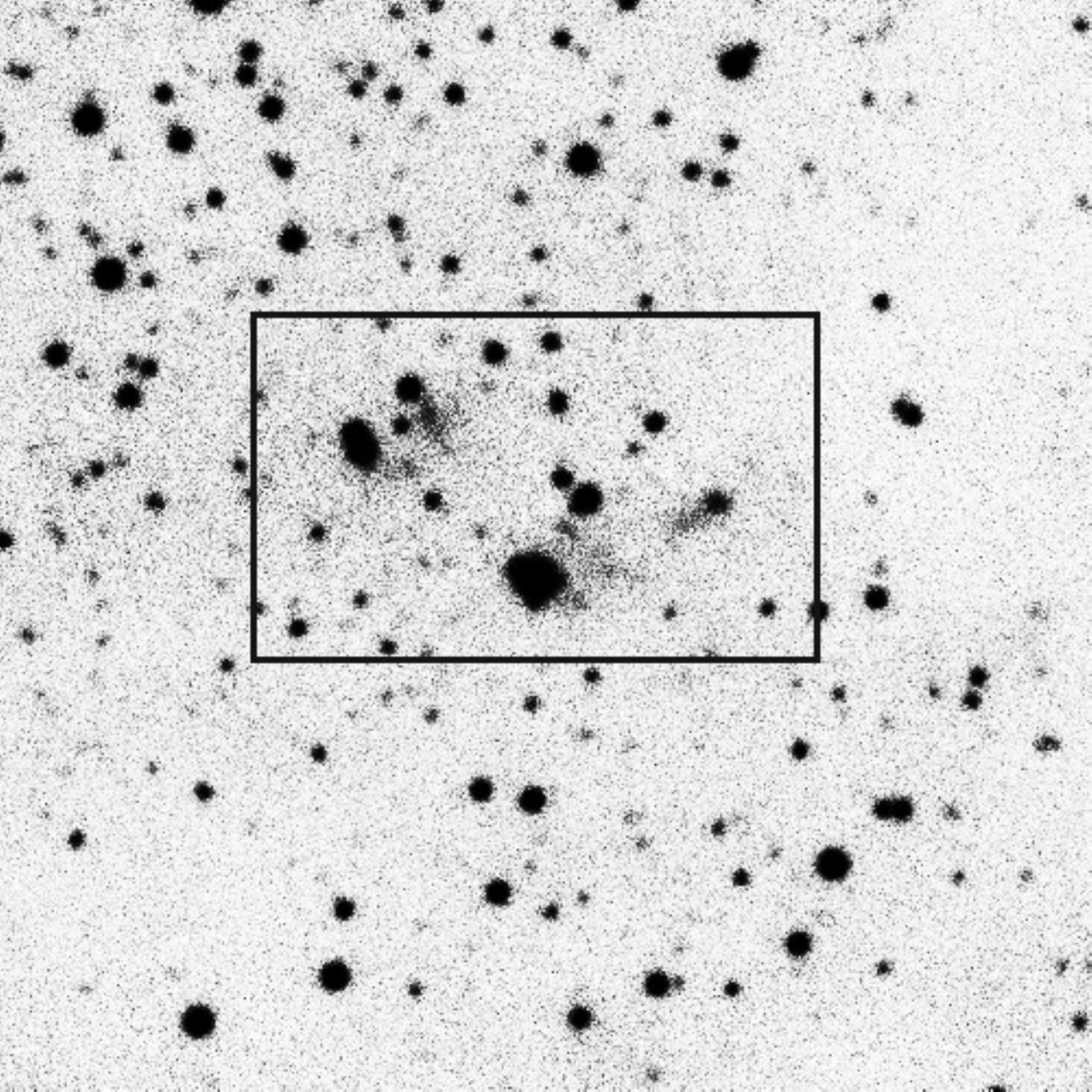}{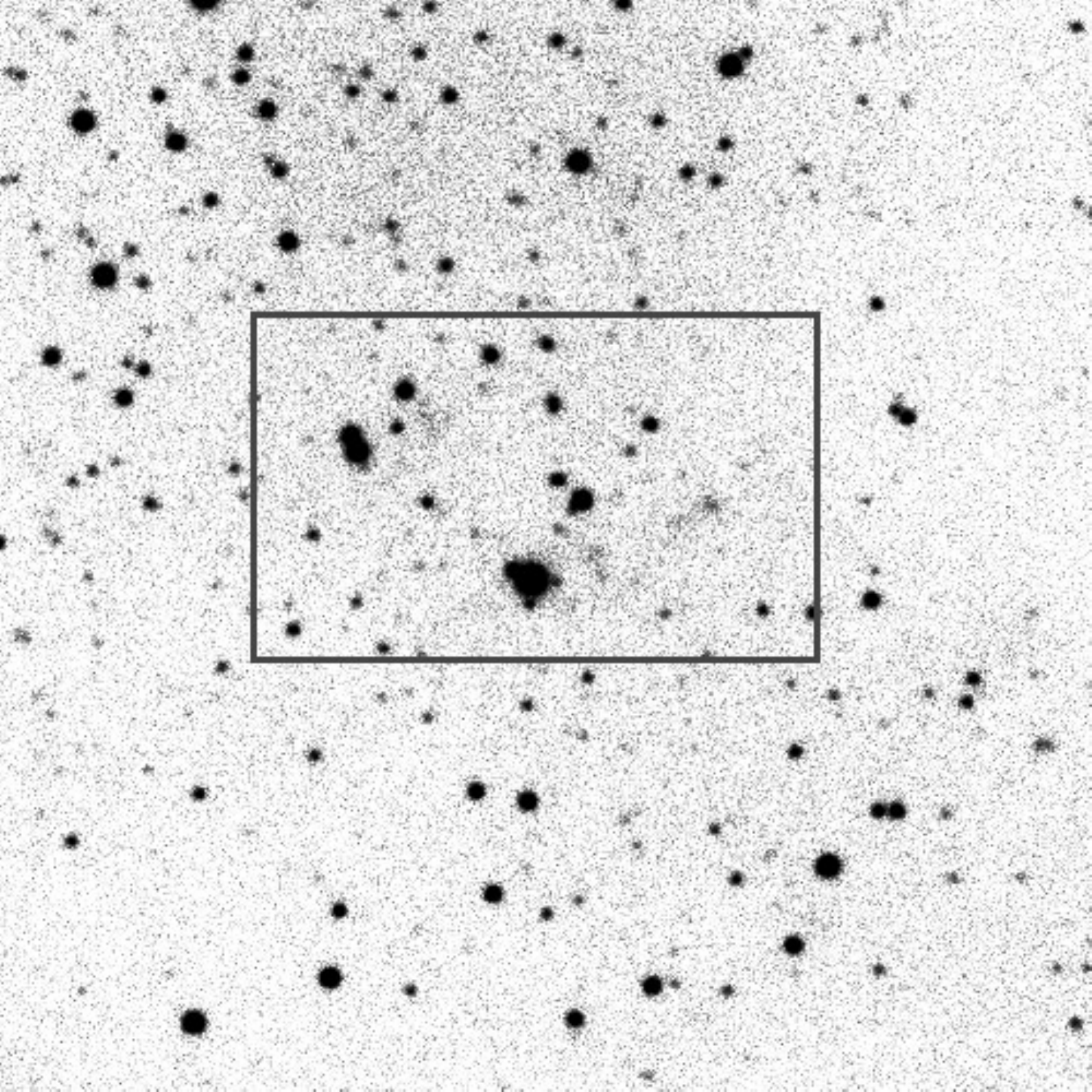}
\caption{HH~1190--1191 in \ha\ (left) and \stwo\ (right).  The field of view for both images is 2\farcm2 square.  Both HH objects are detected in both filters.  The box is the same region as the left box in Figure~\ref{fig:hh567color}.
\label{fig:hh67has2}}
\end{figure}

\begin{figure}[ht]
\plotone{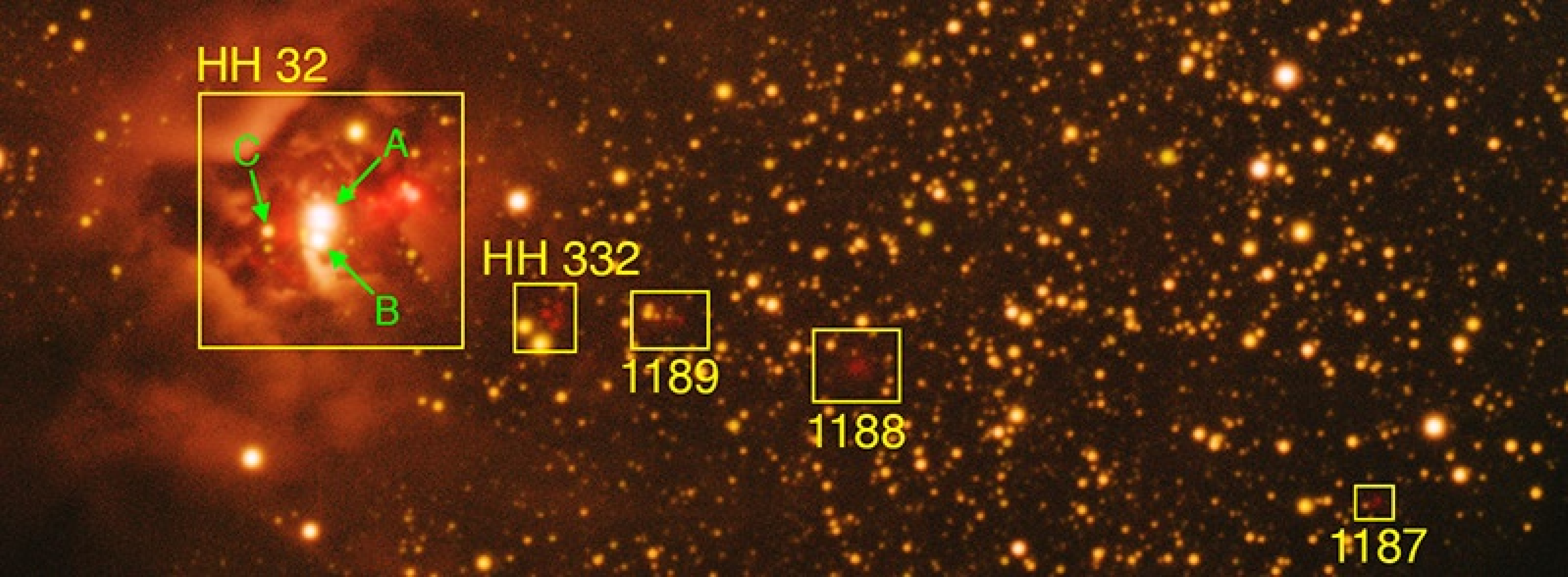}
\plotone{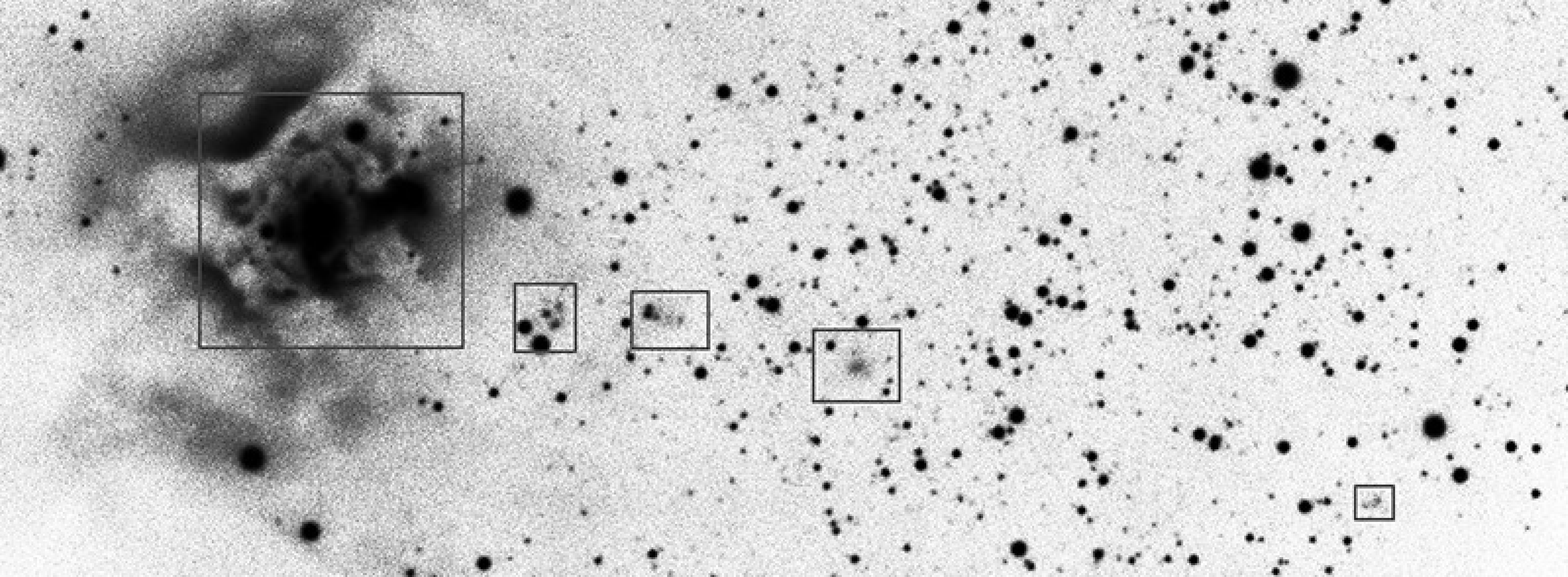}
\plotone{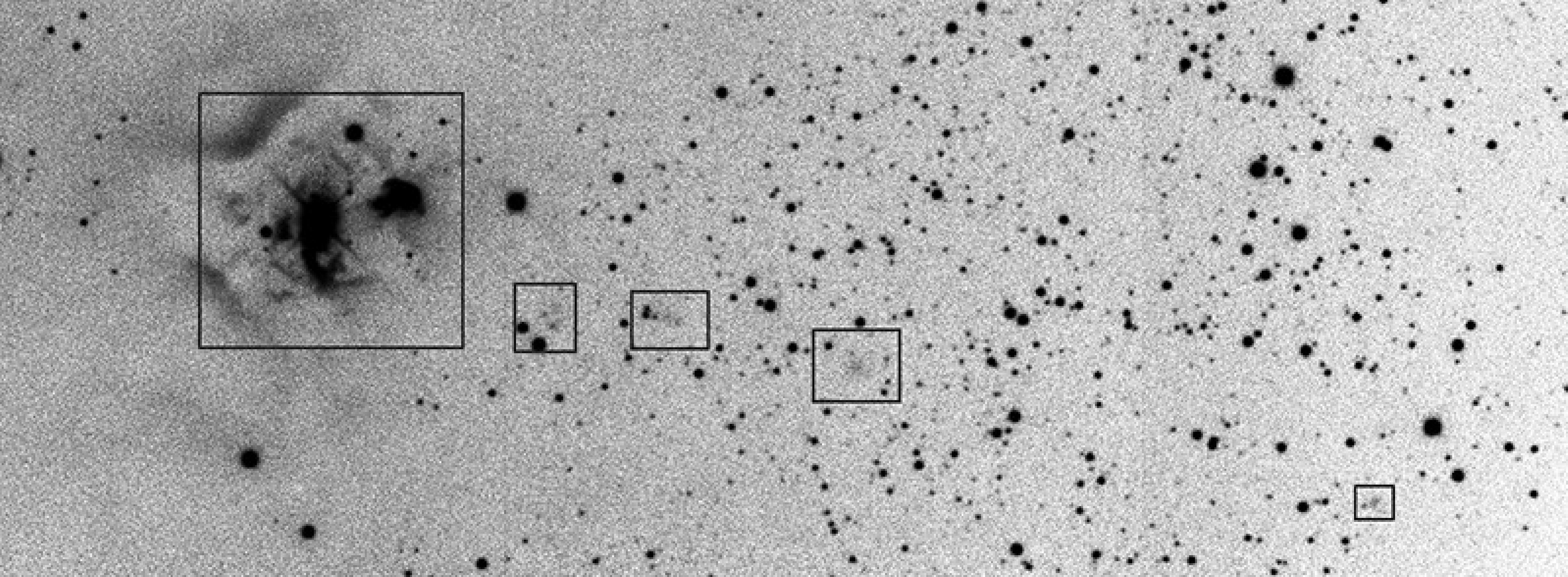}
\caption{The color (top), \ha\ (middle), and \stwo\ (bottom) images of HH~1187--1189 along with HH~32 and HH~332.  The IR sources of AS~353 are labeled in green in the top figure. The field of view is 6\farcm9$\times$2\farcm5.  In the top image the colors for the top frame are the same as for Figure~\ref{fig:hh1-4color}.  
\label{fig:hh89}}
\end{figure}

\begin{figure}[ht]
\plotone{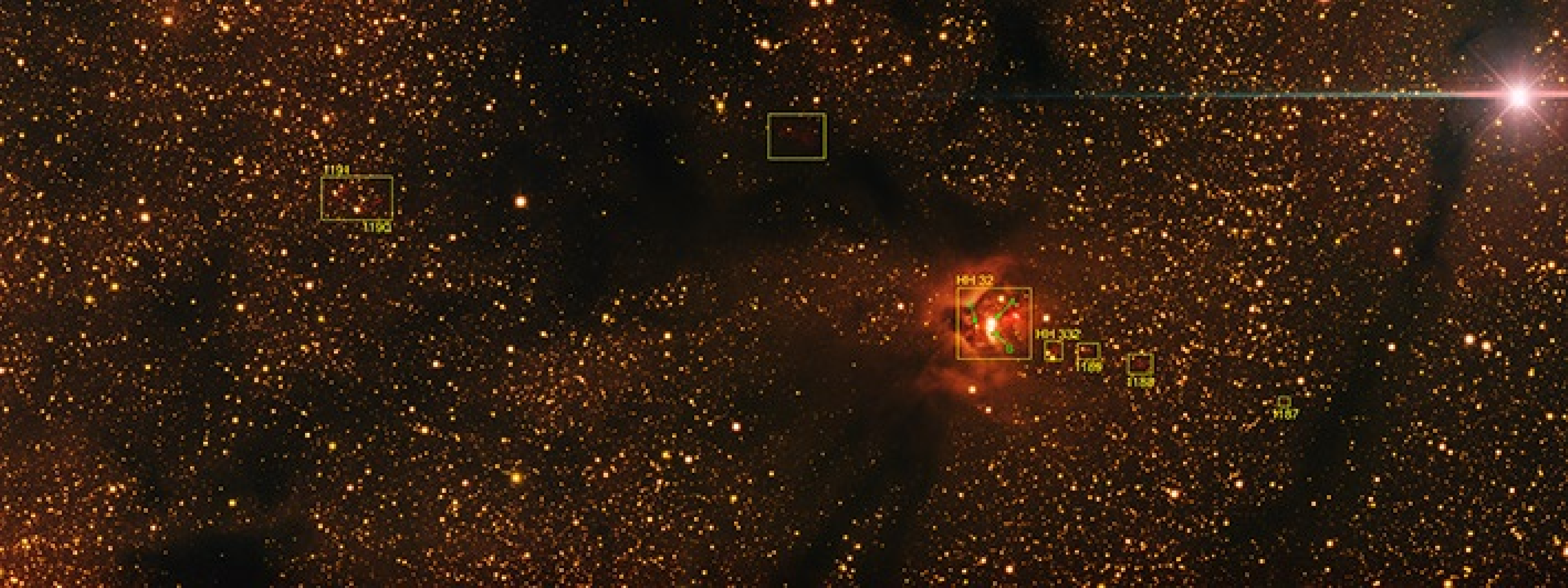}
\plotone{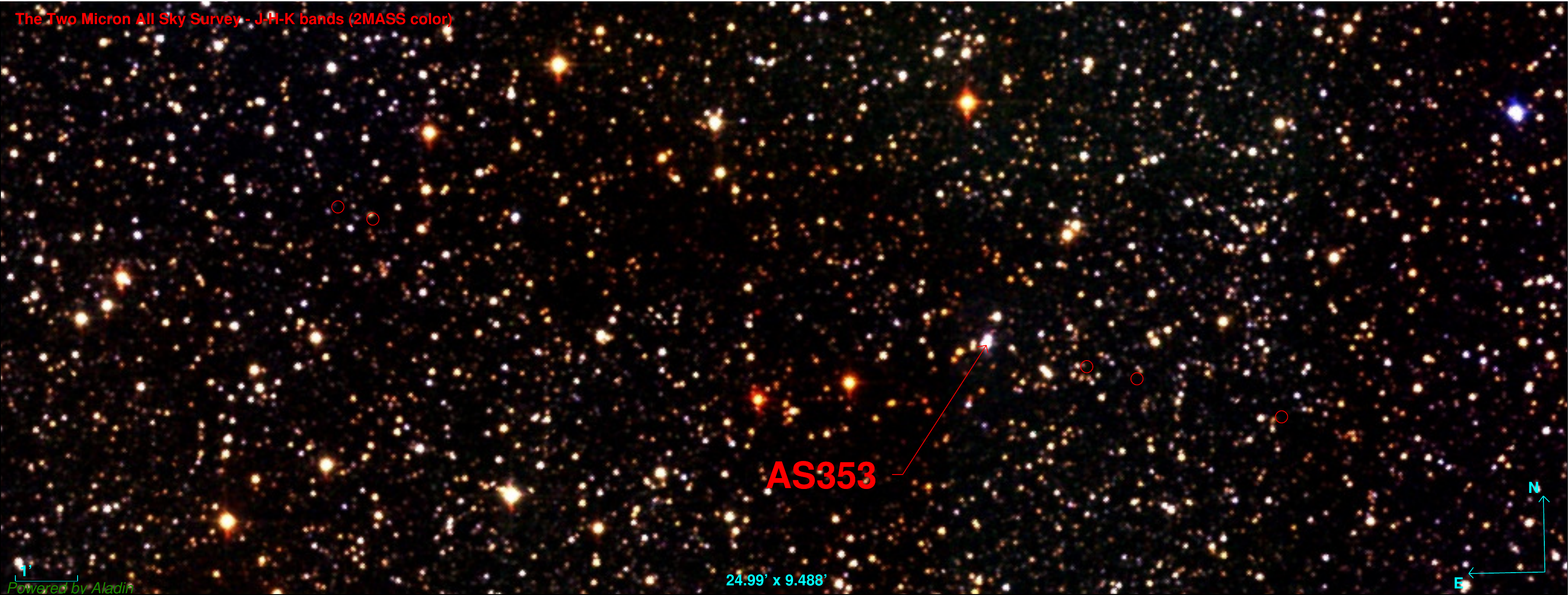}
\plotone{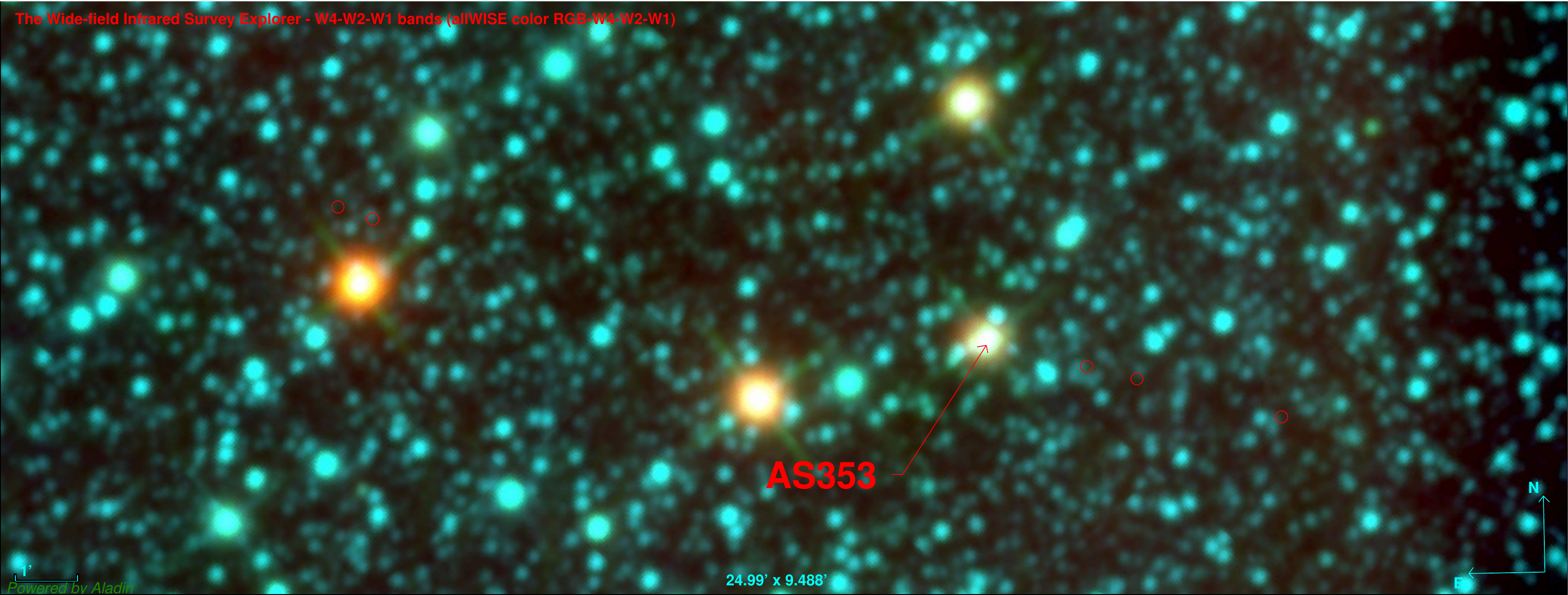}
\caption{The optical (top), 2MASS (middle), and AllWISE (bottom) images of the entire HH~1187--1191 outflow.  The field of view is 25\farcm0$\times$9\farcm5.  In the top frame the colors are the same as for Figure~\ref{fig:hh1-4color}. In the middle frame the J, H, K band are assigned colors of blue, green, and red respectively.  And in the bottom frame, the W1, W2, and W4 bands are assigned to blue, green, and red respectively. Red circles in the middle and bottom images show the locations of the five new HH objects, which are colinear with AS353/HH~32.  The angular distance between HH~1188 and HH~1191 is 15\farcm34 which, at our adopted distance of $200\pm30$~pc for L673 \citep{2006ApJ...647..432R}, corresponds to a projected distance of $0.9\pm0.15$~pc.  The bright WISE source south of HH~1188 and 1189 is an OH/IR star most likely unrelated to the HH objects.
\label{fig:hh69}}
\end{figure}

HH~1187--1191 are quite faint objects and their optical morphologies do not indicate a direction to their origin.  However the locations of these sources are colinear with HH~32 and HH~332, suggesting they all have the same progenitor.  The color-composite and narrowband images for the eastern outflow are shown in Figures~\ref{fig:hh567color} and \ref{fig:hh67has2}.  The western portion of the outflow is shown in Figure~\ref{fig:hh89}.  The full extent of the outflow is shown in optical and IR in Figure~\ref{fig:hh69}.  The coordinates for HH~332 given in \citet{1996ApJ...463..246D} are incorrect. Their positions are about 50" off from those measured in our images.  Updated coordinates are given in Table~\ref{tbl:hh332}.

%This region was not observed with Spitzer, however the WISE and 2MASS surveys reveal three sources a few arcmin south of HH~6/7. The SEDs for these sources are unusual in that there is a large jump between the 2MASS fluxes and the WISE fluxes and the 2MASS fluxes are pretty flat.  In addition, one of the fainter sources in the WISE catalog is resolved into seven 2MASS sources.  It is therefore unlikely these objects are YSOs (BUT WHAT ARE THEY THEN?). 
%UPDATE (RYS):  Turns out the source south of HH 6/7 is an OH/IR star.  I think we can confidently remove these IR sources from the list and discussion.  

HH~32 and 332 are associated with the known young stars AS~353 A and B \citep[for a review see][]{2008hsf1.book...18P}.  If HH~1187--1191 are extensions of the HH~32 and 332 flows, then they are likely also associated with AS~353.  AS353 A and B were both detected in the c2d survey, as well as a fainter source to the east we have labeled AS~353~C (Figure 7).  Both AS353 A and B are identified as ``star+dust'' objects in the c2d catalog, which is consistent with previous spectroscopic work identifying them as T~Tauri stars~\citep{2008hsf1.book...18P}.  Though HH~32 and 332 have been attributed to AS~353 A and B in the past, it is {\em possible} that AS~353~C is also a driving source.  

%NOt sure we need the 2MASS and WISE panels...recommend we use a spitzer IRAC image here that has same FOV roughly as Fig 7 above...

\subsection{HH~1192-1194}

\begin{figure}[ht]
\plotone{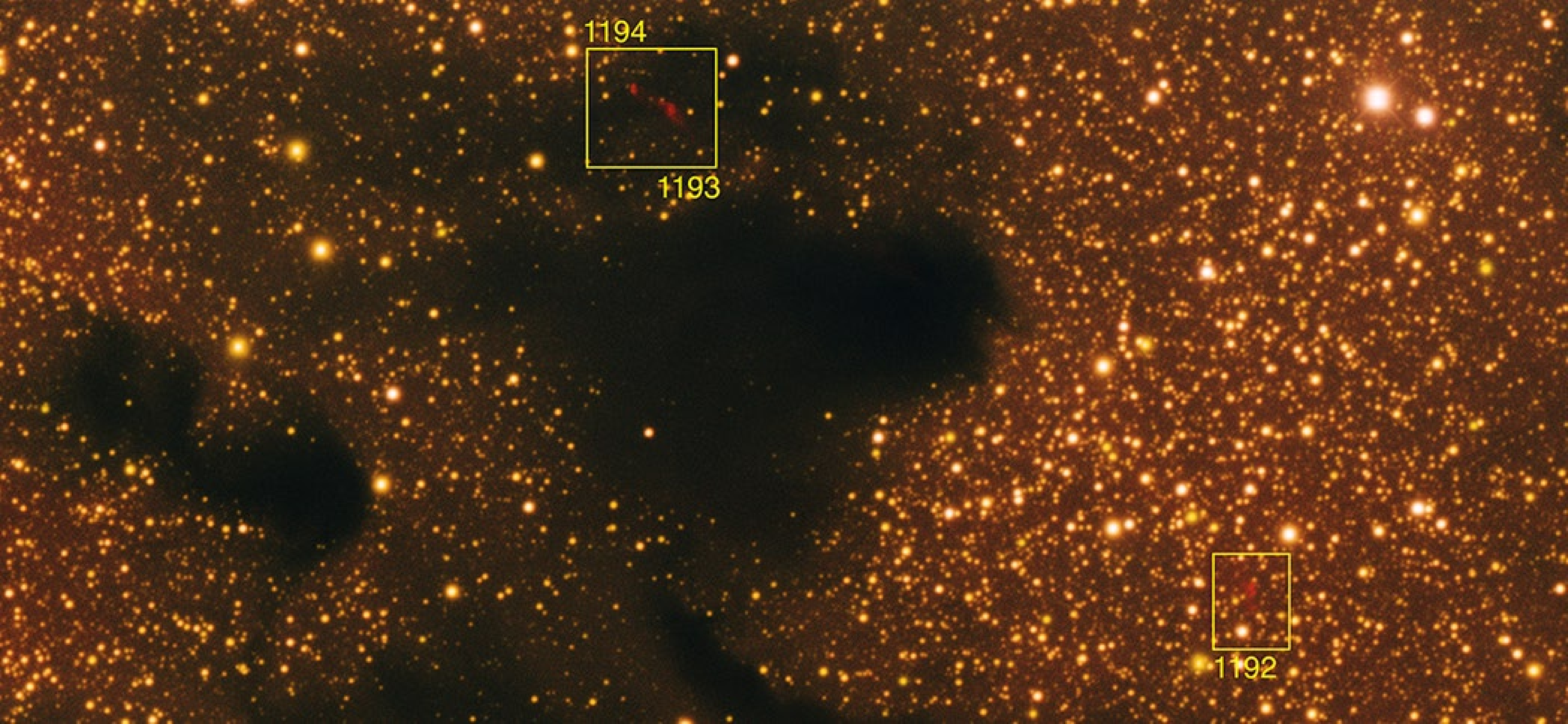}
\caption{The color image of HH~1192-1194.  The field of view is 10\farcm9$\times$5\farcm0.  The colors are the same as for Figure~\ref{fig:hh1-4color}.  
\label{fig:hh1517color}}
\end{figure}

\begin{figure}[ht]
\plottwo{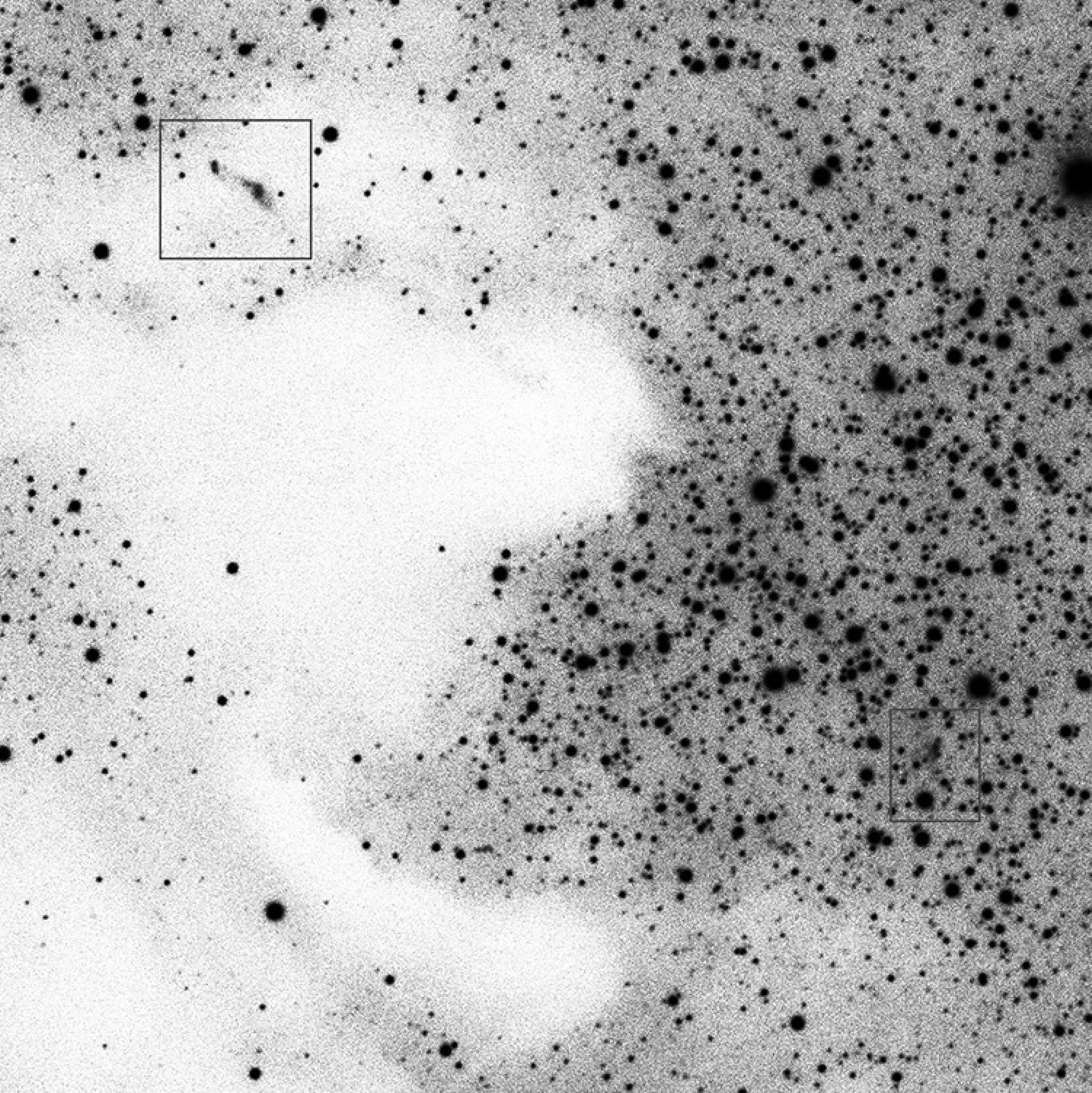}{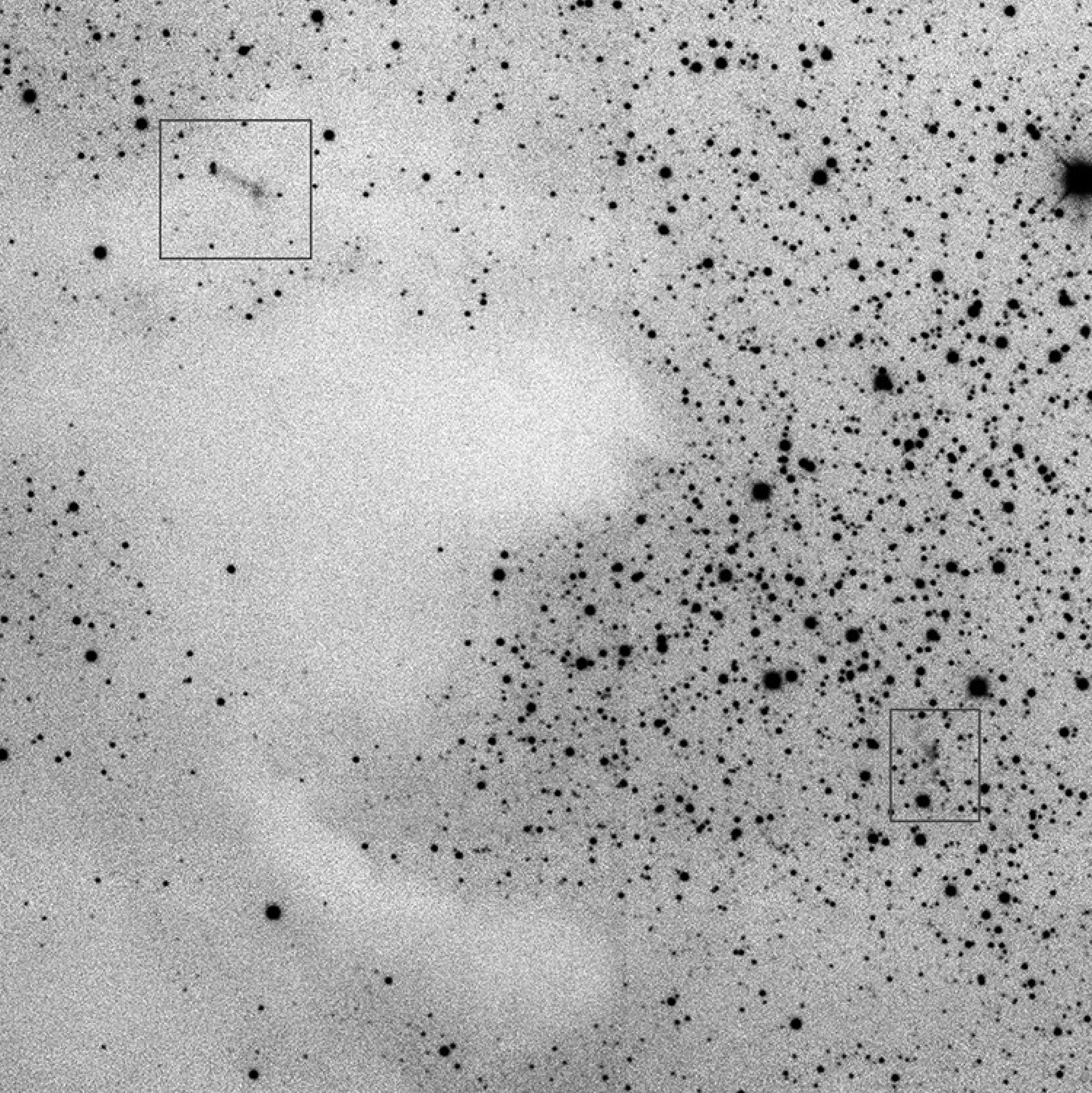}
\caption{HH~1192-1194 in \ha\ (left) and \stwo\ (right).  The field of view for both images is 6\farcm5 square.  All three objects are detected in both filters.
\label{fig:hh1517has2}}
\end{figure}

\begin{figure}[ht]
\plotone{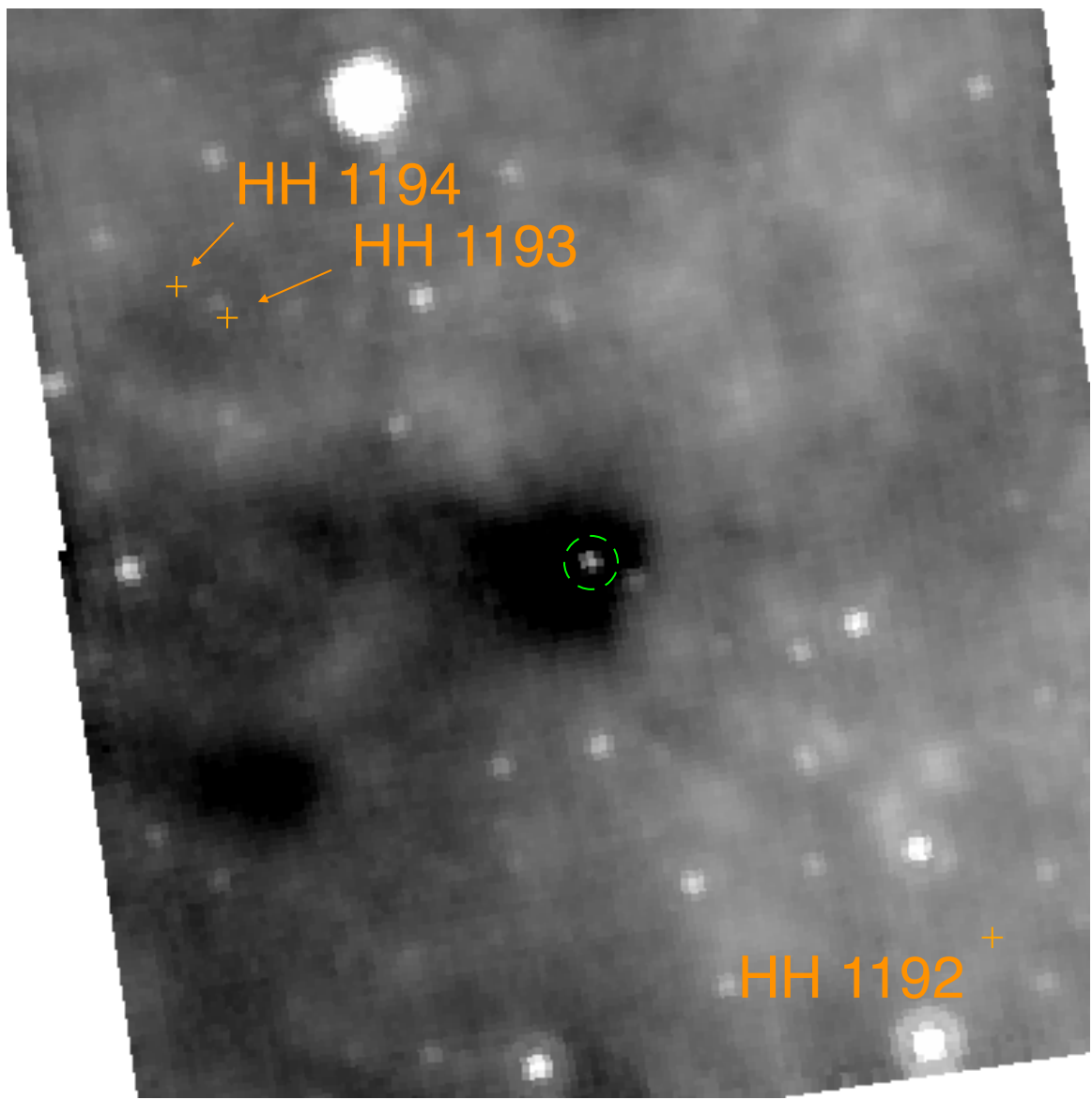}
\caption{{\it Spitzer}-MIPS 24~\micron\ image of the HH~1192-1194 region.  HH object positions are shown in orange and the candidate Class 0 infrared source circled in green.  Note the surrounding infrared dark cloud.  
\label{fig:hh15-17ir}}
\end{figure}

The optical images presented in Figures~\ref{fig:hh1517color} and \ref{fig:hh1517has2} show three HH objects near a dark cloud.  Although there are no obvious bright, red sources in the 2MASS and WISE surveys near HH~1192-1194, there is a ``red'' source in the c2d catalog consistent with a deeply embedded Class 0/I source.  It is not detected in 2MASS, but it is very bright in the MIPS 25 and 70~\micron\ bands, and coincides with an infrared dark cloud (IRDC) (Figure~\ref{fig:hh15-17ir}).  It bisects HH~1192 and HH~1193/1194 and is therefore a potential driver.  It seems likely that this is a deeply embedded Class 0/I source which may be associated with the starless core \object{L673-7} identified by \citet{2004ApJS..152...81P}.  

%replace with MIPS/IRAC image?  And perhaps the WISE band 3/4 image to show the IRDC?
%\begin{figure}[ht]
%\plottwo{HH~11835-17_2MASS.eps}{HH~11835-17_WISE.eps}

%\caption{The 2MASS (left) and AllWISE (right) images of HH~11835--17.  The colors are the same as for Figure~\ref{fig:hh1-4ir}. The small circles show the locations of the three HH objects.  The large (hatched?) circle shows the location of the very red source L673-7.
%\label{fig:hh15-17ir}}
%\end{figure}

\begin{deluxetable}{lll}
\tablecaption{Updated Coordinates for HH~332\label{tbl:hh332}}
\tablewidth{0pt}
\tablehead{\colhead{ID} & \colhead{RA(2000)} & \colhead{DEC}}
\startdata
HH~332A & 19:20:26.7 & +11:01:27 \\
HH~332B & 19:20:26.7 & +11:01:32 \\
HH~332C & 12:20:26.9 & +11:01:30 \\
\enddata
\end{deluxetable}

%Figures~\ref{fig-2} and \ref{fig-3} show the H$\alpha$ and \stwo\ images for each object.

\section{Conclusions}

Our identification of twelve new HH objects provides further evidence of new and on-going star formation in the Aquila region, as HH objects are tracers of the powerful outflows common to the youngest generation of young stars.  This suggests an exciting potential for the future of star formation studies in the Aquila region as these discoveries imply the existence of a larger T~Tauri population than has been previously recognized.  Examination of broad survey data in the region, such as 2MASS, UKIDDS, and WISE, and detailed analyses of color data may reveal more extensive young star populations.  There is particular appeal to the possibility of a substantial young star population in Aquila as the region is less than 300~pc away, providing a nearby population of young stars available for unprecedentedly sensitive and detailed study in the ALMA/JWST era.

\acknowledgments

This work is based in part on observations made with the {\it Spitzer} Space Telescope, which is operated by the Jet Propulsion Laboratory, California Institute of Technology under a contract with NASA. This research has made use of the NASA/ IPAC Infrared Science Archive, which is operated by the Jet Propulsion Laboratory, California Institute of Technology, under contract with the National Aeronautics and Space Administration.
We are grateful to B. Reipurth for many helpful discussions.  We also wish to thank Kitt Peak National Observatory and its excellent support staff.  The figures in this paper were created with the help of the ESA/ESO/NASA FITS Liberator.  This paper is dedicated to the memory of Dr. Katherine Sirles.


\facilities{KPNO:Mayall, Spitzer, WISE, 2MASS, IRSA}.

\bibliography{rector}

\begin{thebibliography}{}
\expandafter\ifx\csname natexlab\endcsname\relax\def\natexlab#1{#1}\fi

\bibitem[{{Andre} {et~al.}(1993){Andre}, {Ward-Thompson}, \&
  {Barsony}}]{1993ApJ...406..122A}
{Andre}, P., {Ward-Thompson}, D., \& {Barsony}, M. 1993, \apj, 406, 122

\bibitem[{{Bacciotti} \& {Eisl{\"o}ffel}(1999)}]{1999A&A...342..717B}
{Bacciotti}, F., \& {Eisl{\"o}ffel}, J. 1999, \aap, 342, 717

\bibitem[{{Bally} {et~al.}(1996){Bally}, {Devine}, \&
  {Reipurth}}]{1996ApJ...473L..49B}
{Bally}, J., {Devine}, D., \& {Reipurth}, B. 1996, \apjl, 473, L49

\bibitem[{{Bally} \& {Reipurth}(2001)}]{2001ApJ...546..299B}
{Bally}, J., \& {Reipurth}, B. 2001, \apj, 546, 299

\bibitem[{{Bontemps} {et~al.}(2010){Bontemps}, {Andr{\'e}}, {K{\"o}nyves},
  {Men'shchikov}, {Schneider}, {Maury}, {Peretto}, {Arzoumanian}, {Attard},
  {Motte}, {Minier}, {Didelon}, {Saraceno}, {Abergel}, {Baluteau}, {Bernard},
  {Cambr{\'e}sy}, {Cox}, {di Francesco}, {di Giorgo}, {Griffin}, {Hargrave},
  {Huang}, {Kirk}, {Li}, {Martin}, {Mer{\'{\i}}n}, {Molinari}, {Olofsson},
  {Pezzuto}, {Prusti}, {Roussel}, {Russeil}, {Sauvage}, {Sibthorpe},
  {Spinoglio}, {Testi}, {Vavrek}, {Ward-Thompson}, {White}, {Wilson},
  {Woodcraft}, \& {Zavagno}}]{2010A&A...518L..85B}
{Bontemps}, S., {Andr{\'e}}, P., {K{\"o}nyves}, V., {et~al.} 2010, \aap, 518,
  L85

\bibitem[{{Dame} {et~al.}(1987){Dame}, {Ungerechts}, {Cohen}, {de Geus},
  {Grenier}, {May}, {Murphy}, {Nyman}, \& {Thaddeus}}]{1987ApJ...322..706D}
{Dame}, T.~M., {Ungerechts}, H., {Cohen}, R.~S., {et~al.} 1987, \apj, 322, 706

\bibitem[{{Davis} {et~al.}(1996){Davis}, {Eisloeffel}, \&
  {Smith}}]{1996ApJ...463..246D}
{Davis}, C.~J., {Eisloeffel}, J., \& {Smith}, M.~D. 1996, \apj, 463, 246

\bibitem[{{Evans} {et~al.}(2003){Evans}, {Allen}, {Blake}, {Boogert}, {Bourke},
  {Harvey}, {Kessler}, {Koerner}, {Lee}, {Mundy}, {Myers}, {Padgett},
  {Pontoppidan}, {Sargent}, {Stapelfeldt}, {van Dishoeck}, {Young}, \&
  {Young}}]{2003PASP..115..965E}
{Evans}, II, N.~J., {Allen}, L.~E., {Blake}, G.~A., {et~al.} 2003, \pasp, 115,
  965

\bibitem[{{Evans} {et~al.}(2009){Evans}, {Dunham}, {J{\o}rgensen}, {Enoch},
  {Mer{\'{\i}}n}, {van Dishoeck}, {Alcal{\'a}}, {Myers}, {Stapelfeldt},
  {Huard}, {Allen}, {Harvey}, {van Kempen}, {Blake}, {Koerner}, {Mundy},
  {Padgett}, \& {Sargent}}]{2009ApJS..181..321E}
{Evans}, II, N.~J., {Dunham}, M.~M., {J{\o}rgensen}, J.~K., {et~al.} 2009,
  \apjs, 181, 321

\bibitem[{{Fazio} {et~al.}(2004){Fazio}, {Hora}, {Allen}, {Ashby}, {Barmby},
  {Deutsch}, {Huang}, {Kleiner}, {Marengo}, {Megeath}, {Melnick}, {Pahre},
  {Patten}, {Polizotti}, {Smith}, {Taylor}, {Wang}, {Willner}, {Hoffmann},
  {Pipher}, {Forrest}, {McMurty}, {McCreight}, {McKelvey}, {McMurray}, {Koch},
  {Moseley}, {Arendt}, {Mentzell}, {Marx}, {Losch}, {Mayman}, {Eichhorn},
  {Krebs}, {Jhabvala}, {Gezari}, {Fixsen}, {Flores}, {Shakoorzadeh}, {Jungo},
  {Hakun}, {Workman}, {Karpati}, {Kichak}, {Whitley}, {Mann}, {Tollestrup},
  {Eisenhardt}, {Stern}, {Gorjian}, {Bhattacharya}, {Carey}, {Nelson},
  {Glaccum}, {Lacy}, {Lowrance}, {Laine}, {Reach}, {Stauffer}, {Surace},
  {Wilson}, {Wright}, {Hoffman}, {Domingo}, \& {Cohen}}]{2004ApJS..154...10F}
{Fazio}, G.~G., {Hora}, J.~L., {Allen}, L.~E., {et~al.} 2004, \apjs, 154, 10

\bibitem[{{Gutermuth} {et~al.}(2008){Gutermuth}, {Bourke}, {Allen}, {Myers},
  {Megeath}, {Matthews}, {J{\o}rgensen}, {Di Francesco}, {Ward-Thompson},
  {Huard}, {Brooke}, {Dunham}, {Cieza}, {Harvey}, \&
  {Chapman}}]{2008ApJ...673L.151G}
{Gutermuth}, R.~A., {Bourke}, T.~L., {Allen}, L.~E., {et~al.} 2008, \apjl, 673,
  L151

\bibitem[{{K{\"o}nyves} {et~al.}(2010){K{\"o}nyves}, {Andr{\'e}},
  {Men'shchikov}, {Schneider}, {Arzoumanian}, {Bontemps}, {Attard}, {Motte},
  {Didelon}, {Maury}, {Abergel}, {Ali}, {Baluteau}, {Bernard}, {Cambr{\'e}sy},
  {Cox}, {di Francesco}, {di Giorgio}, {Griffin}, {Hargrave}, {Huang}, {Kirk},
  {Li}, {Martin}, {Minier}, {Molinari}, {Olofsson}, {Pezzuto}, {Russeil},
  {Roussel}, {Saraceno}, {Sauvage}, {Sibthorpe}, {Spinoglio}, {Testi},
  {Ward-Thompson}, {White}, {Wilson}, {Woodcraft}, \&
  {Zavagno}}]{2010A&A...518L.106K}
{K{\"o}nyves}, V., {Andr{\'e}}, P., {Men'shchikov}, A., {et~al.} 2010, \aap,
  518, L106

\bibitem[{{Lada}(1987)}]{1987IAUS..115....1L}
{Lada}, C.~J. 1987, in IAU Symposium, Vol. 115, Star Forming Regions, ed.
  M.~{Peimbert} \& J.~{Jugaku}, 1--17

\bibitem[{{Monet} {et~al.}(2003){Monet}, {Levine}, {Canzian}, {Ables}, {Bird},
  {Dahn}, {Guetter}, {Harris}, {Henden}, {Leggett}, {Levison}, {Luginbuhl},
  {Martini}, {Monet}, {Munn}, {Pier}, {Rhodes}, {Riepe}, {Sell}, {Stone},
  {Vrba}, {Walker}, {Westerhout}, {Brucato}, {Reid}, {Schoening}, {Hartley},
  {Read}, \& {Tritton}}]{2003AJ....125..984M}
{Monet}, D.~G., {Levine}, S.~E., {Canzian}, B., {et~al.} 2003, \aj, 125, 984

\bibitem[{{Park} {et~al.}(2004){Park}, {Lee}, \& {Myers}}]{2004ApJS..152...81P}
{Park}, Y.-S., {Lee}, C.~W., \& {Myers}, P.~C. 2004, \apjs, 152, 81

\bibitem[{{Prato} {et~al.}(2008){Prato}, {Rice}, \&
  {Dame}}]{2008hsf1.book...18P}
{Prato}, L., {Rice}, E.~L., \& {Dame}, T.~M. 2008, {Where are all the Young
  Stars in Aquila?}, ed. B.~{Reipurth}, 18

\bibitem[{{Rector} {et~al.}(2007){Rector}, {Levay}, {Frattare}, {English}, \&
  {Pu'uohau-Pummill}}]{2007AJ....133..598R}
{Rector}, T.~A., {Levay}, Z.~G., {Frattare}, L.~M., {English}, J., \&
  {Pu'uohau-Pummill}, K. 2007, \aj, 133, 598

\bibitem[{{Rice} {et~al.}(2006){Rice}, {Prato}, \&
  {McLean}}]{2006ApJ...647..432R}
{Rice}, E.~L., {Prato}, L., \& {McLean}, I.~S. 2006, \apj, 647, 432

\bibitem[{{Rieke} {et~al.}(2004){Rieke}, {Young}, {Engelbracht}, {Kelly},
  {Low}, {Haller}, {Beeman}, {Gordon}, {Stansberry}, {Misselt}, {Cadien},
  {Morrison}, {Rivlis}, {Latter}, {Noriega-Crespo}, {Padgett}, {Stapelfeldt},
  {Hines}, {Egami}, {Muzerolle}, {Alonso-Herrero}, {Blaylock}, {Dole}, {Hinz},
  {Le Floc'h}, {Papovich}, {P{\'e}rez-Gonz{\'a}lez}, {Smith}, {Su}, {Bennett},
  {Frayer}, {Henderson}, {Lu}, {Masci}, {Pesenson}, {Rebull}, {Rho}, {Keene},
  {Stolovy}, {Wachter}, {Wheaton}, {Werner}, \&
  {Richards}}]{2004ApJS..154...25R}
{Rieke}, G.~H., {Young}, E.~T., {Engelbracht}, C.~W., {et~al.} 2004, \apjs,
  154, 25

\bibitem[{{Tokunaga} {et~al.}(2004){Tokunaga}, {Reipurth}, {G{\"a}ssler},
  {Hayano}, {Hayashi}, {Iye}, {Kanzawa}, {Kobayashi}, {Kamata}, {Minowa},
  {Nedachi}, {Oya}, {Pyo}, {Saint-Jacques}, {Terada}, {Takami}, \&
  {Takato}}]{2004AJ....127..444T}
{Tokunaga}, A.~T., {Reipurth}, B., {G{\"a}ssler}, W., {et~al.} 2004, \aj, 127,
  444

\bibitem[{{Werner} {et~al.}(2004){Werner}, {Roellig}, {Low}, {Rieke}, {Rieke},
  {Hoffmann}, {Young}, {Houck}, {Brandl}, {Fazio}, {Hora}, {Gehrz}, {Helou},
  {Soifer}, {Stauffer}, {Keene}, {Eisenhardt}, {Gallagher}, {Gautier}, {Irace},
  {Lawrence}, {Simmons}, {Van Cleve}, {Jura}, {Wright}, \&
  {Cruikshank}}]{2004ApJS..154....1W}
{Werner}, M.~W., {Roellig}, T.~L., {Low}, F.~J., {et~al.} 2004, \apjs, 154, 1

\bibitem[{{White} {et~al.}(2002){White}, {Hillenbrand}, {Metchev}, \&
  {Patience}}]{2002AAS...201.2005W}
{White}, R.~J., {Hillenbrand}, L., {Metchev}, S., \& {Patience}, J. 2002, in
  Bulletin of the American Astronomical Society, Vol.~34, American Astronomical
  Society Meeting Abstracts, 1134

\end{thebibliography}

\end{document}